\documentclass[preprint,aps,prb,showpacs,mathserif, superscriptaddress, preprintnumbers]{revtex4-1}
\usepackage{times}
\usepackage{graphicx}
\usepackage{amsmath}
\usepackage{lineno}

\usepackage{natmove}
\usepackage [english]{babel}
\usepackage{float}
\usepackage [autostyle, english = american]{csquotes}
\MakeOuterQuote{"}

\newcommand{\eqn}[2]{\begin{equation}\label{#2}#1\end{equation}}
\newcommand{\mt}[1]{\mathrm{#1}}

\begin{document}



\title{Small-mass atomic defects enhance vibrational thermal transport at disordered interfaces with ultrahigh thermal boundary conductance}
\author{Ashutosh Giri}
\affiliation{Department of Mechanical and Aerospace Engineering, University of Virginia, Charlottesville, Virginia 22904, USA}

\author{Sean W. King}
\affiliation{Intel Corporation, Logic Technology Development, 5200 NE Elam Young Parkway, Hillsboro, Oregon 97124}

\author{William A. Lanford}
\affiliation{Department of Physics, University at Albany, State University of New York, Albany, NY, 12222, USA}

\author{Antonio R. Mei}
\affiliation{Intel Corporation, Logic Technology Development, 5200 NE Elam Young Parkway, Hillsboro, Oregon 97124}

\author{Devin Merril}
\affiliation{Intel Corporation, Logic Technology Development, 5200 NE Elam Young Parkway, Hillsboro, Oregon 97124}

\author{Liza Ross}
\affiliation{Intel Corporation, Logic Technology Development, 5200 NE Elam Young Parkway, Hillsboro, Oregon 97124}

\author{Ron Oviedo}
\affiliation{Intel Corporation, Logic Technology Development, 5200 NE Elam Young Parkway, Hillsboro, Oregon 97124}

\author{John Richards}
\affiliation{Intel Corporation, Logic Technology Development, 5200 NE Elam Young Parkway, Hillsboro, Oregon 97124}

\author{David H. Olson}
\affiliation{Department of Mechanical and Aerospace Engineering, University of Virginia, Charlottesville, Virginia 22904, USA}

\author{Jeffrey L. Braun}
\affiliation{Department of Mechanical and Aerospace Engineering, University of Virginia, Charlottesville, Virginia 22904, USA}

\author{John T. Gaskins}
\affiliation{Department of Mechanical and Aerospace Engineering, University of Virginia, Charlottesville, Virginia 22904, USA}

\author{Freddy DeAngelis}
\affiliation{George W. Woodruff School of Mechanical Engineering, Georgia Institute of Technology, Atlanta, GA 30332, USA}

\author{Asegun Henry}
\affiliation{George W. Woodruff School of Mechanical Engineering, Georgia Institute of Technology, Atlanta, GA 30332, USA}
\affiliation{School of Materials Science and Engineering, Georgia Institute of Technology, Atlanta, GA, 30332, USA}

\author{Patrick E. Hopkins}
\email{phopkins@virginia.edu}
\affiliation{Department of Mechanical and Aerospace Engineering, University of Virginia, Charlottesville, Virginia 22904, USA}
\affiliation{Department of Materials Science and Engineering, University of Virginia, Charlottesville, Virginia 22904, USA}
\affiliation{Department of Physics, University of Virginia, Charlottesville, Virginia 22904, USA}

\date{\today}
\begin{abstract}
The role of interfacial nonidealities and disorder on thermal transport across interfaces is traditionally assumed to add resistance to heat transfer, decreasing the thermal boundary conductance (TBC).\cite{hopkins2013aa} However, recent computational works have suggested that interfacial defects can enhance this thermal boundary conductance through the emergence of unique vibrational modes that are intrinsic to the material interface and defect atoms,\cite{duda2012ac,murakami2014aa,gordiz2016ab,giri2017ad,gordiz2016aa} a finding that contradicts traditional theory and conventional understanding. By manipulating the local heat flux of atomic vibrations that comprise these interfacial modes, in principle, the TBC can be increased. In this work, we provide evidence that interfacial defects can enhance the TBC across interfaces through the emergence of unique high frequency vibrational modes that arise from atomic mass defects at the interface with relatively small masses. We demonstrate ultrahigh TBC at amorphous SiOC:H/SiC:H interfaces, approaching 1 GW m$^{-2}$ K$^{-1}$, that is further increased through the introduction of nitrogen defects. The fact that disordered interfaces can exhibit such high conductances, which can be further increased with additional defects, offers a unique direction to manipulate heat transfer across materials with high densities of interfaces by controlling and enhancing interfacial thermal transport. 


\end{abstract}

\maketitle

\newpage
Heterogeneous interfaces between two adjacent solids have enabled the realization of ultralow thermal conductivity materials,\cite{chiritescu2007aa,costescu2004aa,pernot2010aa,losego2013aa,hopkins2011af} with reduction to thermal conductivity often falling below the corresponding minimum limit traditionally attributed to a pure amorphous solid.\cite{einstein1911aa} These reductions in thermal conductivity have been driven by additional temperature drops occurring at each interface. These temperature drops are quantified by the TBC, which is traditionally assumed to be related to the phonon states in each material comprising the interface. Thus, using the approach to engineer materials with high densities of interfaces to achieve ultralow thermal conductivity solids requires a fundamental understanding of how atomic vibrations interact and exchange energy at interfaces, which, with the advent of disorder and other nanoscale features, is arguably lacking. 

While including disorder in a crystalline system can lead to reductions in thermal conductivity, this same phenomena may not hold true at interfaces. Recent theories have suggested that vibrational modes unique to the interfaces that do not exist intrinsically in any of the homogeneous materials can in fact contribute substantially to the TBC.\cite{saaskilahti2014aa,chalopin2013aa,duda2011ab,murakami2014aa,gordiz2016ab,giri2017ad,gordiz2016aa,giri2016af} Therefore, judiciously selected defects near the interface could in principle be used to increase the TBC by enhancing these interfacial modes. In this case, disordered interfaces could lead to higher TBCs than more `perfect' interfaces. Indeed, recent computational works have demonstrated that the TBC at amorphous/amorphous and amorphous/crystalline interfaces can be higher than that at crystalline/crystalline interfaces comprised of the same material.\cite{gordiz2017aa,giri2015ad,giri2016ac} This reasoning can not be explained by conventional phonon TBC theories,\cite{swartz1989aa,reddy2005aa,hopkins2009af} and offers a unique picture of how vibrational energy couples across defected/disordered interfaces. However, experimental demonstrations of the existence of these interfacial defect modes and their contributions to TBC are lacking. 

In this work, we report on the thermal conductivity of a series of amorphous multilayers (AMLs) comprised of alternating layers of hydrogenated amorphous silicon carbide (a-SiC:H) and hydrogenated amorphous silicon oxycarbide (a-SiOC:H) with varying interface densities. As the heat transport in these AMLs is completely diffusive, we extract a TBC across the a-SiC:H/a-SiOC:H interface that approaches 1 GW m$^{-2}$ K$^{-1}$, the highest diffusive TBC measured to date. Through an {\it{in situ}} plasma exposure, we introduce N$_2$ defects at and near the interface in each layer of the AML. The introduction of these defects causes the thermal conductivity of these AMLs to be independent of interface density; in other words, the resistance at the interfaces disappears, or the TBC increases beyond the ability to measure a quantifiable value. Supported with both vibrational spectroscopy and molecular dynamics simulations, we identify interfacial defect modes that arise in the thermal phonon regime only in the N$_2$-processed AMLs.  

\begin{figure*}
\begin{center}
\includegraphics[width=6in]{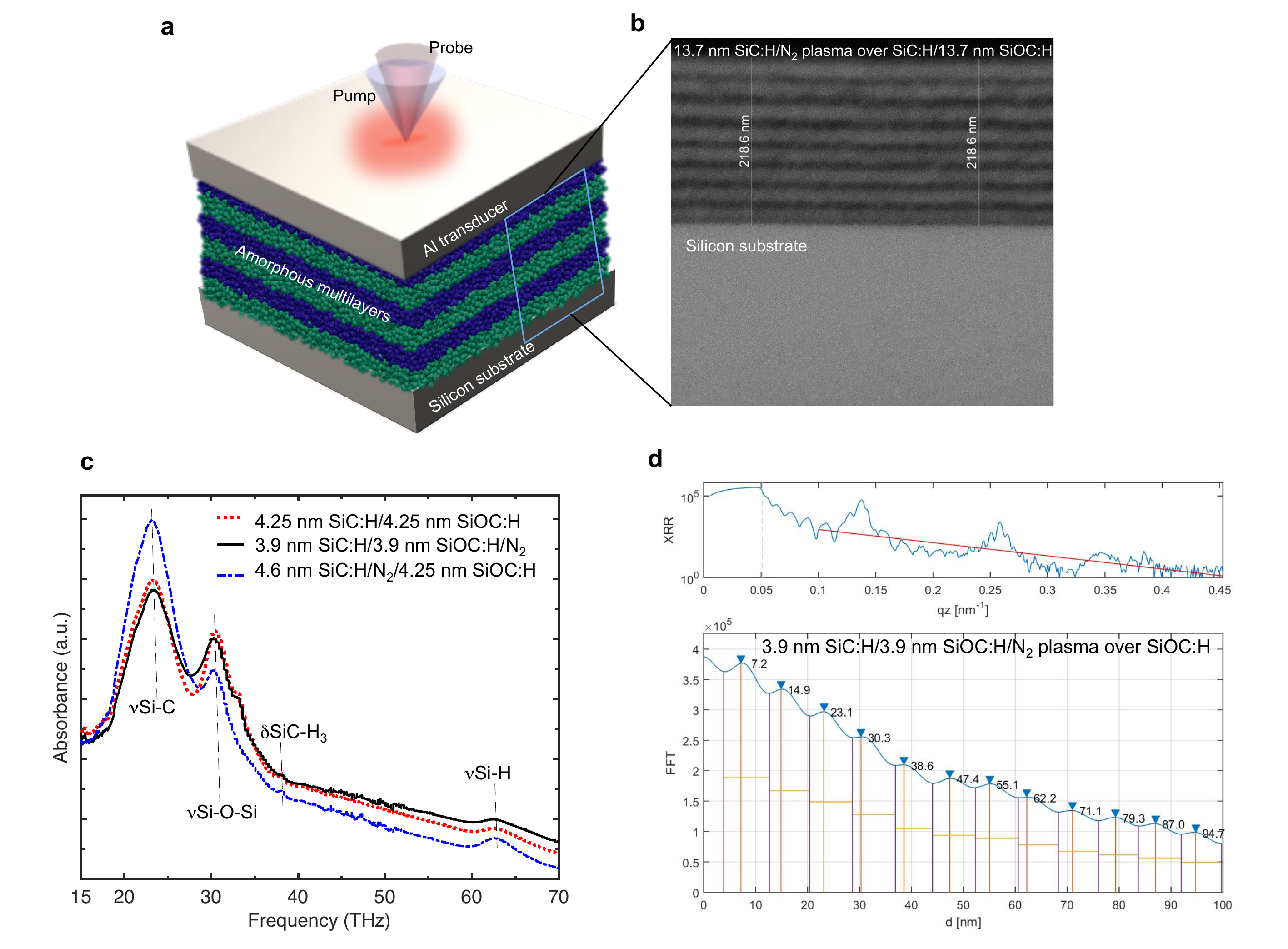}
\caption{(a) Schematic of a multilayer sample for our thermal measurements via the pump-probe TDTR technique. (b) Characteristic XSEM image for a multilayer with 27.4 nm period thickness and N$_2$ plasma treatment carried out on the surface of SiC:H layers. The thickness and periodicity can be confirmed via the XSEM images. (c) Characteristic FTIR spectra for three representative samples with and without N$_2$ plasma treatment on the SiOC:H or SiC:H laminates {\it{in situ}} during growth. (d) Characteristic XRR patterns showing multilayer reflections exemplified by the peaks in the XRR data for a (7.8 nm period thick) SiC:H/SiOC:H with N$_2$ plasma treatment over SiOC:H layers. (Bottom Panel) The FFT analysis on the XRR confirms the periodicity of the multilayers. }
\label{Characterization}

\end{center}
\end{figure*}

The amorphous SiOC:H/SiC:H multilayer samples were deposited on crystalline silicon substrates via plasma-enhanced chemical vapor deposition (PECVD). A sample series with N$_2$ plasma treated multilayers, carried out {\it{in situ}} during growth between the deposition of either the SiOC:H or the SiC:H layers, were also fabricated to understand the effect of interfacial nonidealities that arise due to lighter atoms at the interface on mediating thermal transport across disordered interfaces. A schematic of the sample used for our thermal measurements is shown in Fig.~\ref{Characterization}a. The film and period thicknesses were determined via x-ray reflectivity (XRR) and cross-section scanning electron microscopy (XSEM) measurements; example XSEM and XRR measurements are shown in Fig.~\ref{Characterization}b and Fig.~\ref{Characterization}d, for a SiOC:H/SiC:H multilayer with N$_2$ plasma treatments carried out on the surface of the SiOCH layers, and for a multilayer with N$_2$ plasma treatments carried out on the surface of the SiC:H layers, respectively. The chemical compositions of the multilayer films and homogeneous samples were determined via rutherford backscattering (RBS) spectroscopy (details in the methods section). The percent composition of C, N, O, Si and H are tabulated in Table S1 of the Supplementary Information. Along with the chemical compositions, the density of the films were calculated via RBS and are tabulated in Table S1 of the Supplementary Information. The vibrational properties of the samples were studied using Fourier transform infrared (FTIR) spectroscopy.  Figure~\ref{Characterization}c shows example spectra for SiOC:H/SiC:H multilayers with and without N$_2$ plasma. In comparison to the SiOC:H/SiC:H sample, the similarities between the FTIR spectra of the sample in which the SiC:H was treated with N$_2$ vs. the sample in which the SiOC:H was treated with plasma suggests the N$_2$ plasma is enhancing the vibrations in the 20-30 THz range, as discussed in detail later.

To measure the thermal properties, we employed the time domain thermoreflectance (TDTR) technique, which is a non-contact optical pump-probe technique. First, we measure the thermal conductivity and heat capacity of individual SiOC:H and SiC:H films as a function of film thickness (as shown in Supplementary Fig.~S7). The lack of film thickness dependence on the thermal conductivity for the  a-SiOC:H and a-SiC:H films suggests that heat conduction is mostly driven by vibrations that are non-propagating (e.g., diffusons and locons).\cite{larkin2014aa} This is in contrast to our recent experimental results demonstrating size effects on the thermal conductivity of amorphous Si thin films,\cite{braun2016aa} where a significant portion of heat flow is due to propagons that represent delocalized propagating modes. The lack of size effects in the thermal conductivity of a-SiOC:H can be attributed to the Si-O-Si network structure confirmed from the FTIR measurements (Supplementary Fig.~S3a), which is similar to the structure found in SiO$_2$; the lack of size effects in SiO$_2$ is primarily due to the weak bonding that exists between the SiO$_4$ tetrahedra, whereas, the thickness dependent thermal conductivity in a-Si is a result of strongly bonded tetrahedra.\cite{larkin2014aa} For the a-SiC:H films, the FTIR results (as shown in Supplementary Fig.~S3b) show that the network structure mostly consists of Si-C stretching modes similar to a-SiC systems;\cite{chen2016ab,king2011aa} the lack of size effects in the a-SiC:H is consistent with size independent thermal conductivities measured for amorphous SiC in Ref.~\onlinecite{jeong2012aa}. These findings along with the measurement of heat capacities for the amorphous SiOC:H and SiC:H films and the measured thermal conductivities of amorphous SiOC:H/SiC:H SLs with varying period thicknesses are used to derive a mean TBC across a single SiOC:H/SiC:H interface as detailed in the discussions below.

\begin{figure}
\begin{center}
\includegraphics[width=3.25in]{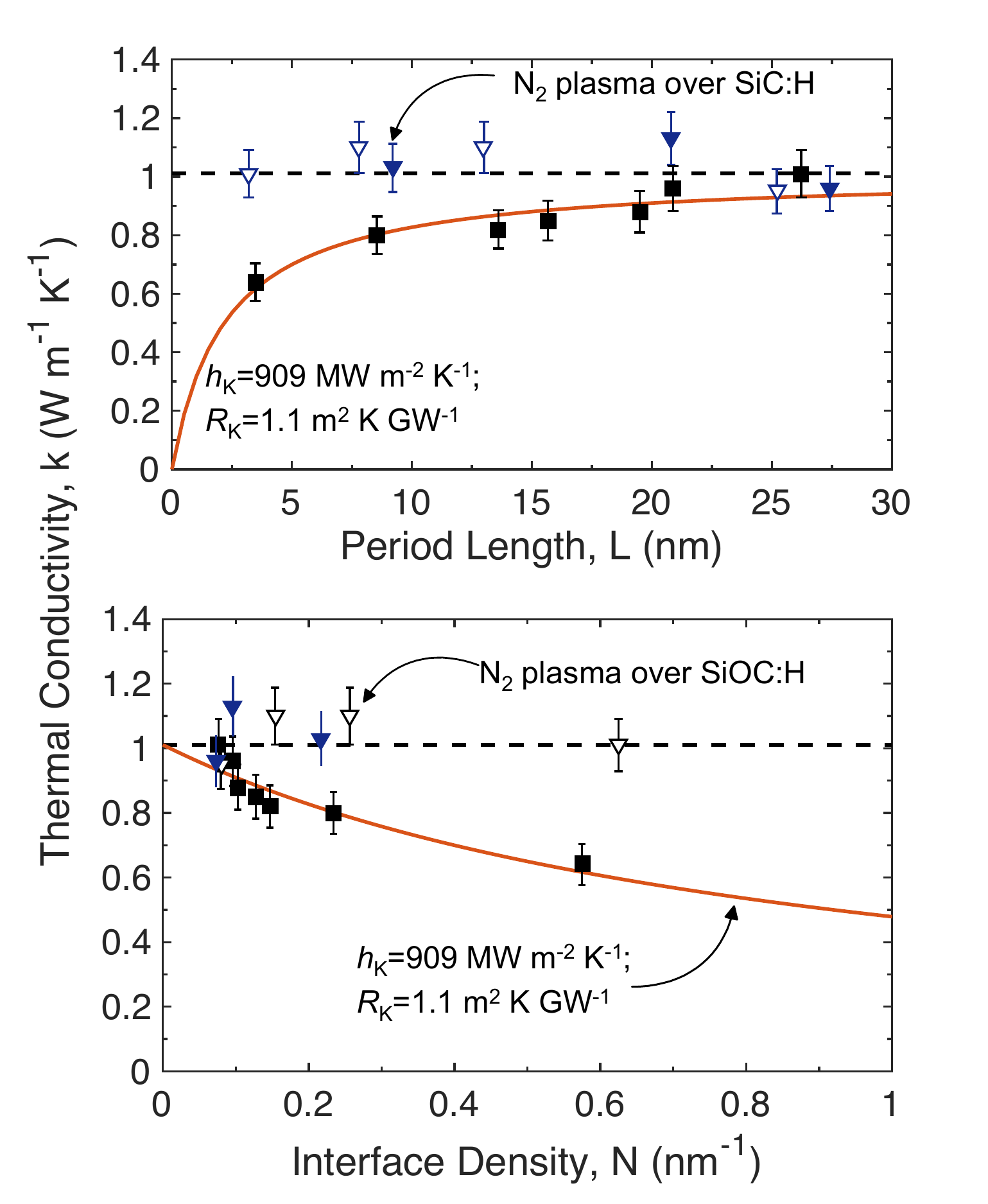}
\caption{Thermal conductivities of amorphous SiOC:H/SiC:H superlattices plotted as a function of (a) period length and (b) interface density. For comparison, thermal conductivities of N$_2$ plasma treated superlattices are also included. The N$_2$ plasma is shown to increase the thermal conductivities of AMLs with smaller period thicknesses.}
\label{TCvsL_N}
\end{center}
\end{figure}


\begin{figure*}
\begin{center}
\includegraphics[width=6.75in]{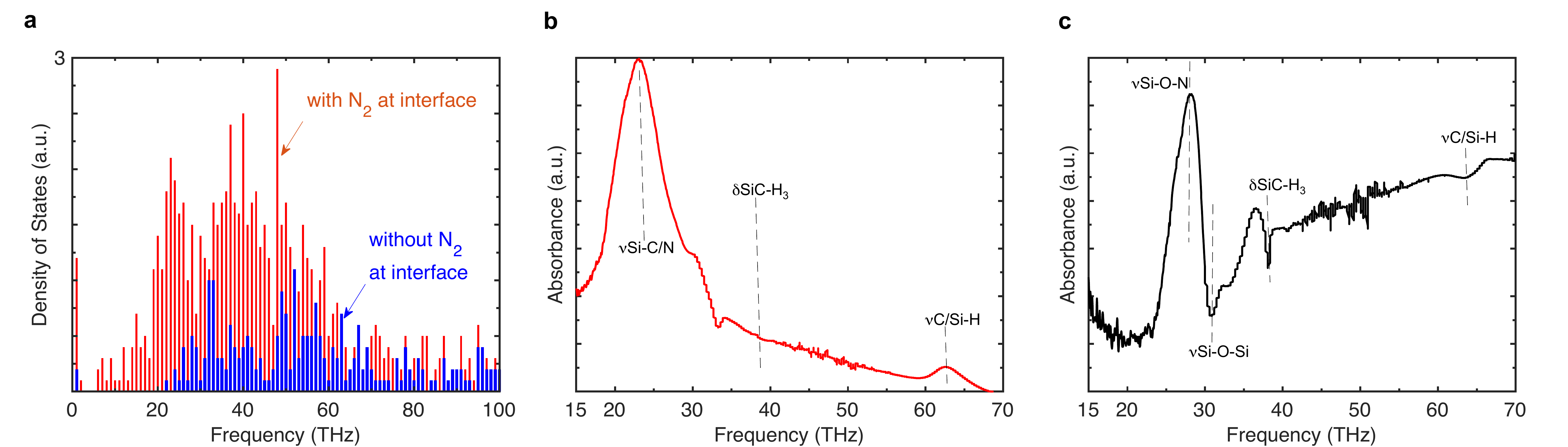}
\caption{(a) MD-predicted interfacial density of states for our amorphous multilayers with and without N$_2$ atoms at the interface. Subtracted FTIR spectra for SiC:H/SiOC:H multilayers with N$_2$ plasma treatment over (b) SiC:H and (c) SiOC:H with respect to a multilayer without the plasma treatment.}
\label{DOSfig}
\end{center}
\end{figure*}

The measured thermal conductivities of the amorphous SiOC:H/SiC:H superlattices are shown as a function of period lengths and interface densities in Fig.~\ref{TCvsL_N}a and Fig.~\ref{TCvsL_N}b, respectively (square symbols). The thermal conductivity for SiC:H/SiOC:H SLs monotonically decreases with decreasing period thickness and increasing interface density. This suggests that the interfaces in the amorphous SLs contribute non-negligibly to thermal resistance across the thin films. To determine the TBC across the SiC:H/SiOC:H interface, we apply the widely used thermal circuit model,\cite{hopkins2012ab,giri2015ad} which describes the resistivity, $\rho$, of a SL as a superposition of the thermal resistances of the individual layers and the resistances at the individual interfaces as,
\eqn{\rho=\kappa^{-1}=\frac{1}{L}\bigg[\frac{L}{2\kappa_\mt{SiOC:H}}+\frac{L}{2\kappa_\mt{SiC:H}}+2R_\mt{K}\bigg].} 
HHere $\kappa_\mt{SiOC:H}$ and $\kappa_\mt{SiC:H}$ are determined from the measurements of the thickness series for the respective homogeneous samples. Equation~1 is fit to the experimental data with $R_\mt{K}$ as the fitting parameter. Using this approach, we determine $R_\mt{K}$=1.1 m$^{2}$ K GW$^{-1}$, alternatively the TBC, $h_\mt{K}$=1/$R_\mt{K}$=909 MW m$^{-2}$ K$^{-1}$), resulting in the best fit line shown in Fig.~\ref{TCvsL_N}. Note, this TBC is considerably higher than nearly all TBCs reported in the literature for crystalline/crystalline interfaces (typical TBCs at crystalline/crystalline interfaces range from $\sim$20 to 300 MW m$^{-2}$ K$^{-1}$).\cite{hopkins2013aa}

The high TBCs at these amorphous SiC:H/SiOC:H interfaces are in line with those predicted via molecular dynamics simulations  (Refs.~\onlinecite{giri2015ac,gordiz2017aa}), experimentally measured across SiO$_2$/Al$_2$O$_3$ interfaces reported from a single AML at room temperature (0.67 GW m$^{-2}$ K$^{-1}$),\cite{fong2016aa} and the lower limit to TBC measured across an amorphous SiO$_2$/crystalline Si interface.\cite{kimling2017aa} In Ref.~\onlinecite{giri2015ac}, we showed that the TBC across a generic Lennard Jones (LJ)-based amorphous/amorphous interface is higher than that of their crystalline counterpart, suggesting that TBC associated with amorphous interfaces are, in general, much higher than those across their corresponding crystalline interfaces. An analysis to predict the spectral contributions at the LJ-based amorphous/amorphous and crystalline/crystalline interfaces (as detailed in the Supplementary Information) suggests that vibrations carrying heat across interfaces are very different between the amorphous and crystalline phases. Along these lines, recent work has suggested that disorder around amorphous interfaces force the atomic vibrations near the interface to perturb the natural modes of vibrations in the amorphous materials, leading to higher frequency vibrations near the interface that effectively couple with each other.\cite{gordiz2017aa} Thus, in the event that the masses of these atoms are reduced, the local velocity that drives the cross-correlation of the heat flux will be increased. In this regard, the introduction of light atom impurities at amorphous interfaces should further increase the TBC by enabling a higher heat flux across the interface. 

Figure~\ref{TCvsL_N} shows the thermal conductivity of the multilayers with N$_2$ plasma treatment carried out after either the SiOC:H or SiC:H layers are deposited. For both cases, when N$_2$ plasma is exposed on the SiOC:H layers or on the SiC:H layers, the thermal conductivity of the multilayers are independent of period thicknesses, in contrast to the results for the multilayers without the plasma treatment. Note, only the sample with the thinnest period with N$_2$ plasma treatment over SiOC:H layers does not demonstrate superlattice type peaks in the XRR. Thus, the thermal conductivity measurement for this sample marks the disordered `alloy limit' for the samples with the N$_2$ plasma treatment. 

As shown in Table S1 of the Supplementary Information, the chemical compositions and the density of the multilayers do not change significantly due to the plasma treatment, which suggests that the varying thermal conductivity trends as shown in Fig.~\ref{TCvsL_N} for our AMLs with/without plasma treatments is not due to densification or drastic changes in the composition and coordination number for these films. Furthermore, the thermal conductivity of samples with plasma treatment carried out at different thickness intervals for homogeneous SiC:H and SiOC:H films do not change within uncertainty compared to the ones without the plasma treatment (Table S1 of Supplementary Information). These observations suggest that there is a different mechanism leading to an increase in the thermal conductivity of the N$_2$ plasma treated samples at high interface densities. To investigate this phenomena further, we turn to material specific molecular dynamics simulations for our structures. Figure~\ref{DOSfig}a shows the density of states (DOS) for the interfacial modes predicted by a super cell lattice dynamics (SCLD) calculation for a short 2.5 nm period AML structure. The SCLD calculations used the ReaxFF to model the interatomic interactions\cite{duin2001aa} and the definition of an `interfacial mode' was taken to be the same as what was used previously by Gordiz and Henry.\cite{gordiz2016ab} Here, the interfacial region was taken to be all atoms within $\pm$7 $\mt{\AA}$ of the interface. As expected, since the two different systems contain different atom types in the interfacial region, the structures with and without nitrogen atoms at the interface exhibit differences in the interfacial modes that manifest. Most notably, there is a substantial increase (2X) in the total fraction of interfacial modes when the nitrogen is introduced. It should be noted that the total bulk DOS calculated did not show a significant change overall, but there is a significant change in the fraction of modes that are localized near the interface (increasing from 1.09 to 4.02\%). This is to be expected, since new and uniquely tailored solutions to the equations of motion are required for the nitrogen atoms, which differ from the atoms everywhere else in the structure. 

This is further validated by our FTIR measurements on the AMLs, shown in Figs.~\ref{DOSfig}b and \ref{DOSfig}c, where we plot the absorbance for the samples with the plasma treatment on the SiC:H and SiOC:H layers, respectively, which is subtracted from the absorbance for the multilayer with similar period thickness without the plasma treatments. There is some correspondence between the FTIR results and the changes in interfacial mode DOS. Notably, the most significant differences between the interfacial DOS in both cases arise in the 20-40 THz regime, where the highest FTIR absorption is observed. Furthermore, the subtracted FTIR results in Figs~\ref{DOSfig}b and ~\ref{DOSfig}c, show that the most significant differences occur in the same frequency interval 20-40 THz. For the multilayer with N$_2$ exposed to the SiC:H layers, there is a clear increase in the vibrational bands at $\sim$24 THz that are associated with Si-C and Si-N bonds. For the case where we ran the N2 plasma on top of the SiOC:H, we clearly see the appearance of an Si-O/N mode at $\sim$27 THz. This corresponds with a decrease in absorbance for the Si-O-Si stretching mode, SiC-H$_3$ deformation mode, and C-H stretching mode as shown by the dips in the absorbance spectra in  Fig.~\ref{DOSfig}c. Taken together, the FTIR results are consistent with our MD-predicted DOS increase for these modes, which arise at the interface. Given that the total DOS for both structures is virtually indistinguishable (as shown in Supplementary information Fig.~S), and the fact that there is a noticeable change in the FTIR results, it suggests that the interfacial modes, which are different for the two structures, may be responsible for the difference in IR absorption. 

From these discussions and observations, we can attribute the increase in thermal conductivity and subsequent removal of the interfacial resistance for the plasma treated samples, as shown in Fig.~\ref{TCvsL_N}, to the incorporation of light atoms at the interfacial region that enhances the heat carrying interfacial modes in our AMLs. These results suggest that the TBC at these already ultrahigh TBC interfaces can increase to values $>$1 GW m$^{-2}$ K$^{-1}$ with the inclusion of nitrogen interfacial defects.

\section*{Acknowledgements}

This work was supported in part by the Office of Naval research, Grant. No. N00014-15-12769.

\section*{Methods}
Mono- and multi-layers of the SiC:H ($k$ = 6.5) and SiOC:H ($k$ = 3.2) dielectrics were deposited on 300 mm diameter Si (001) substrates via plasma enhanced chemical vapor deposition (PECVD) the details of which have been previously provided.\cite{king2011aa,guo2015aa} Briefly, a commercially available PECVD system and tetramethylsilane diluted in hydrogen or carbon dioxide were utilized to deposit at 400 $^{\circ}$C the SiC:H and SiOC:H films, respectively. 

XRR spectra were collected using a Bede Fab200 Plus employing a Cu microbeam source and an asymmetric cut Ge crystal. The data was collected in the range of 0 to 9000-15000 arcseconds with approximately 20 arcsecond steps.\cite{king2012ab} Spectra were acquired and fitted using the REFS software package (version 4.0 Bede). Along with XRR, the thickness of the films, multi-layers, and individual laminates were also determined by XSEM measurements performed using an FEI Helios Nanolab scanning electron microscope at magnifications of 20-100,000X.\cite{miikkulainen2015aa}  Note, the XRR and XSEM data confirm the periodic arrangement of SiOC:H/SiC:H layers, with period thicknesses ranging from $\sim$3.4 to 26.4 nm. 

Transmission FTIR spectra were collected from the SiCH and SiOCH films and multilayers using a Thermo Scientific Nicolet 6700 FTIR spectrometer and deuterated L-alanine doped triglycine sulfate (DLaTGS) detector. The spectra were acquired from 400--7000 cm$^{-1}$ with 4 cm$^{-1}$  resolution and signal averaged over 256 scans. Absorption from the silicon substrate was removed by scanning an uncoated silicon substrate and subtracting this spectrum from that for the film (or multilayer)/substrate spectrum. Thin film-substrate optical interference effects were removed using rigorous methods that accounted for the full wave nature of light and have been previously described in detail.\cite{milosevic2012aa,king2012aa,stan2017aa} 

Prior to the thermal measurements, we deposit $\sim$80 nm of Al transducer layer, the thickness of which is measured via picosecond acoustics.\cite{thomsen1986aa} We measure the thermal properties of these samples via TDTR, the details of the technique and the analysis procedure are given elsewhere.\cite{cahill2004aa,schmidt2008aa,hopkins2010aa} Along with the theoretical fits, Figs.~suppinfo shows TDTR data for our homogeneous SiOC:H and SiC:H samples as well as for a multilayer with $\sim$3.48 nm period thickness. To accurately determine the thermal conductivities of the SiOC:H/SiC:H SLs, the heat capacities of the constituent layers in the SL must be determined first. Along with the thermal conductivities of the SLs, to accurately determine the intrinsic resistance at a single SiOC:H/SiC:H interface, the thermal conductivities of the individual layers are also required (as described in detail below). For this purpose, the thermal conductivities and heat capacities of the thickness series of homogeneous a-SiOC:H and a-SiC:H films are determined with a similar approach as implemented in our earlier work in Ref.~\onlinecite{giri2016aa}. In this approach, different pump-modulation frequencies are utilized to simultaneously measure the thermal conductivities and heat capacities of SiC:H and SiOC:H thin films. Figures~suppinfo and~suppinfo show the sensitivities of the ratio to the thermophysical parameters in the thermal model for a $\sim$88 nm thick SiC:H film at 2 and 10 MHz pump-modulation frequencies, respectively. The most sensitive parameters in the model are the heat capacity and the thermal conductivity of the thin film. However, for a particular frequency, a range of heat capacities can produce the best-fit to the experimental data as shown in the contour plot of the residual error (see Fig.~suppinfo). The common heat capacity and thermal conductivity values that produce the best-fit to the data for the two frequencies are taken as the measured values for the homogeneous samples. Using this approach, we measure thermal conductivity values of 1.48$\pm$0.12 W m$^{-1}$ K$^{-1}$ and 0.75$\pm$0.06 W m$^{-1}$ K$^{-1}$ and volumetric heat capacities of 1.9$\pm$0.3 J cm$^{-3}$ K$^{-1}$ and 1.3$\pm$0.2 J cm$^{-3}$ K$^{-1}$ for a-SiC:H and a-SiOC:H, respectively.


\begin{thebibliography}{43}%
\makeatletter
\providecommand \@ifxundefined [1]{%
 \@ifx{#1\undefined}
}%
\providecommand \@ifnum [1]{%
 \ifnum #1\expandafter \@firstoftwo
 \else \expandafter \@secondoftwo
 \fi
}%
\providecommand \@ifx [1]{%
 \ifx #1\expandafter \@firstoftwo
 \else \expandafter \@secondoftwo
 \fi
}%
\providecommand \natexlab [1]{#1}%
\providecommand \enquote  [1]{``#1''}%
\providecommand \bibnamefont  [1]{#1}%
\providecommand \bibfnamefont [1]{#1}%
\providecommand \citenamefont [1]{#1}%
\providecommand \href@noop [0]{\@secondoftwo}%
\providecommand \href [0]{\begingroup \@sanitize@url \@href}%
\providecommand \@href[1]{\@@startlink{#1}\@@href}%
\providecommand \@@href[1]{\endgroup#1\@@endlink}%
\providecommand \@sanitize@url [0]{\catcode `\\12\catcode `\$12\catcode
  `\&12\catcode `\#12\catcode `\^12\catcode `\_12\catcode `\%12\relax}%
\providecommand \@@startlink[1]{}%
\providecommand \@@endlink[0]{}%
\providecommand \url  [0]{\begingroup\@sanitize@url \@url }%
\providecommand \@url [1]{\endgroup\@href {#1}{\urlprefix }}%
\providecommand \urlprefix  [0]{URL }%
\providecommand \Eprint [0]{\href }%
\providecommand \doibase [0]{http://dx.doi.org/}%
\providecommand \selectlanguage [0]{\@gobble}%
\providecommand \bibinfo  [0]{\@secondoftwo}%
\providecommand \bibfield  [0]{\@secondoftwo}%
\providecommand \translation [1]{[#1]}%
\providecommand \BibitemOpen [0]{}%
\providecommand \bibitemStop [0]{}%
\providecommand \bibitemNoStop [0]{.\EOS\space}%
\providecommand \EOS [0]{\spacefactor3000\relax}%
\providecommand \BibitemShut  [1]{\csname bibitem#1\endcsname}%
\let\auto@bib@innerbib\@empty
\bibitem [{\citenamefont {Hopkins}(2013)}]{hopkins2013aa}%
  \BibitemOpen
  \bibfield  {author} {\bibinfo {author} {\bibfnamefont {P.~E.}\ \bibnamefont
  {Hopkins}},\ }\href@noop {} {\bibfield  {journal} {\bibinfo  {journal} {ISRN
  Mechanical Engineering}\ ,\ \bibinfo {pages} {682586}} (\bibinfo {year}
  {2013})}\BibitemShut {NoStop}%
\bibitem [{\citenamefont {Duda}\ \emph {et~al.}(2012)\citenamefont {Duda},
  \citenamefont {English}, \citenamefont {Piekos}, \citenamefont {Beechem},
  \citenamefont {Kenny},\ and\ \citenamefont {Hopkins}}]{duda2012ac}%
  \BibitemOpen
  \bibfield  {author} {\bibinfo {author} {\bibfnamefont {J.~C.}\ \bibnamefont
  {Duda}}, \bibinfo {author} {\bibfnamefont {T.~S.}\ \bibnamefont {English}},
  \bibinfo {author} {\bibfnamefont {E.~S.}\ \bibnamefont {Piekos}}, \bibinfo
  {author} {\bibfnamefont {T.~E.}\ \bibnamefont {Beechem}}, \bibinfo {author}
  {\bibfnamefont {T.~W.}\ \bibnamefont {Kenny}}, \ and\ \bibinfo {author}
  {\bibfnamefont {P.~E.}\ \bibnamefont {Hopkins}},\ }\href {\doibase
  http://dx.doi.org/10.1063/1.4757941} {\bibfield  {journal} {\bibinfo
  {journal} {Journal of Applied Physics}\ }\textbf {\bibinfo {volume} {112}},\
  \bibinfo {eid} {073519} (\bibinfo {year} {2012}),\
  http://dx.doi.org/10.1063/1.4757941}\BibitemShut {NoStop}%
\bibitem [{\citenamefont {Murakami}\ \emph {et~al.}(2014)\citenamefont
  {Murakami}, \citenamefont {Hori}, \citenamefont {Shiga},\ and\ \citenamefont
  {Shiomi}}]{murakami2014aa}%
  \BibitemOpen
  \bibfield  {author} {\bibinfo {author} {\bibfnamefont {T.}~\bibnamefont
  {Murakami}}, \bibinfo {author} {\bibfnamefont {T.}~\bibnamefont {Hori}},
  \bibinfo {author} {\bibfnamefont {T.}~\bibnamefont {Shiga}}, \ and\ \bibinfo
  {author} {\bibfnamefont {J.}~\bibnamefont {Shiomi}},\ }\href
  {http://stacks.iop.org/1882-0786/7/i=12/a=121801} {\bibfield  {journal}
  {\bibinfo  {journal} {Applied Physics Express}\ }\textbf {\bibinfo {volume}
  {7}},\ \bibinfo {pages} {121801} (\bibinfo {year} {2014})}\BibitemShut
  {NoStop}%
\bibitem [{\citenamefont {Gordiz}\ and\ \citenamefont
  {Henry}(2016{\natexlab{a}})}]{gordiz2016ab}%
  \BibitemOpen
  \bibfield  {author} {\bibinfo {author} {\bibfnamefont {K.}~\bibnamefont
  {Gordiz}}\ and\ \bibinfo {author} {\bibfnamefont {A.}~\bibnamefont {Henry}},\
  }\href@noop {} {\bibfield  {journal} {\bibinfo  {journal} {Journal of Applied
  Physics}\ }\textbf {\bibinfo {volume} {119}},\ \bibinfo {eid} {015101}
  (\bibinfo {year} {2016}{\natexlab{a}})}\BibitemShut {NoStop}%
\bibitem [{\citenamefont {Giri}\ and\ \citenamefont
  {Hopkins}(2017)}]{giri2017ad}%
  \BibitemOpen
  \bibfield  {author} {\bibinfo {author} {\bibfnamefont {A.}~\bibnamefont
  {Giri}}\ and\ \bibinfo {author} {\bibfnamefont {P.~E.}\ \bibnamefont
  {Hopkins}},\ }\href {\doibase 10.1038/s41598-017-10482-z} {\bibfield
  {journal} {\bibinfo  {journal} {Scientific Reports}\ }\textbf {\bibinfo
  {volume} {7}},\ \bibinfo {pages} {11011} (\bibinfo {year}
  {2017})}\BibitemShut {NoStop}%
\bibitem [{\citenamefont {Gordiz}\ and\ \citenamefont
  {Henry}(2016{\natexlab{b}})}]{gordiz2016aa}%
  \BibitemOpen
  \bibfield  {author} {\bibinfo {author} {\bibfnamefont {K.}~\bibnamefont
  {Gordiz}}\ and\ \bibinfo {author} {\bibfnamefont {A.}~\bibnamefont {Henry}},\
  }\href@noop {} {\bibfield  {journal} {\bibinfo  {journal} {Scientific
  Reports}\ }\textbf {\bibinfo {volume} {6}},\ \bibinfo {pages} {23139}
  (\bibinfo {year} {2016}{\natexlab{b}})}\BibitemShut {NoStop}%
\bibitem [{\citenamefont {Chiritescu}\ \emph {et~al.}(2007)\citenamefont
  {Chiritescu}, \citenamefont {Cahill}, \citenamefont {Nguyen}, \citenamefont
  {Johnson}, \citenamefont {Bodapati}, \citenamefont {Keblinski},\ and\
  \citenamefont {Zschack}}]{chiritescu2007aa}%
  \BibitemOpen
  \bibfield  {author} {\bibinfo {author} {\bibfnamefont {C.}~\bibnamefont
  {Chiritescu}}, \bibinfo {author} {\bibfnamefont {D.~G.}\ \bibnamefont
  {Cahill}}, \bibinfo {author} {\bibfnamefont {N.}~\bibnamefont {Nguyen}},
  \bibinfo {author} {\bibfnamefont {D.}~\bibnamefont {Johnson}}, \bibinfo
  {author} {\bibfnamefont {A.}~\bibnamefont {Bodapati}}, \bibinfo {author}
  {\bibfnamefont {P.}~\bibnamefont {Keblinski}}, \ and\ \bibinfo {author}
  {\bibfnamefont {P.}~\bibnamefont {Zschack}},\ }\href {\doibase
  10.1126/science.1136494} {\bibfield  {journal} {\bibinfo  {journal}
  {Science}\ }\textbf {\bibinfo {volume} {315}},\ \bibinfo {pages} {351}
  (\bibinfo {year} {2007})}\BibitemShut {NoStop}%
\bibitem [{\citenamefont {Costescu}\ \emph {et~al.}(2004)\citenamefont
  {Costescu}, \citenamefont {Cahill}, \citenamefont {Fabreguette},
  \citenamefont {Sechrist},\ and\ \citenamefont {George}}]{costescu2004aa}%
  \BibitemOpen
  \bibfield  {author} {\bibinfo {author} {\bibfnamefont {R.~M.}\ \bibnamefont
  {Costescu}}, \bibinfo {author} {\bibfnamefont {D.~G.}\ \bibnamefont
  {Cahill}}, \bibinfo {author} {\bibfnamefont {F.~H.}\ \bibnamefont
  {Fabreguette}}, \bibinfo {author} {\bibfnamefont {Z.~A.}\ \bibnamefont
  {Sechrist}}, \ and\ \bibinfo {author} {\bibfnamefont {S.~M.}\ \bibnamefont
  {George}},\ }\href@noop {} {\bibfield  {journal} {\bibinfo  {journal}
  {Science}\ }\textbf {\bibinfo {volume} {303}},\ \bibinfo {pages} {989}
  (\bibinfo {year} {2004})}\BibitemShut {NoStop}%
\bibitem [{\citenamefont {Pernot}\ \emph {et~al.}(2010)\citenamefont {Pernot},
  \citenamefont {Stoffel}, \citenamefont {Savic}, \citenamefont {Pezzoli},
  \citenamefont {Chen}, \citenamefont {Savelli}, \citenamefont {Jacquot},
  \citenamefont {Schumann}, \citenamefont {Denker}, \citenamefont {M{\"o}nch},
  \citenamefont {Deneke}, \citenamefont {Schmidt}, \citenamefont {Rampnoux},
  \citenamefont {Wang}, \citenamefont {Plissonnier}, \citenamefont {Rastelli},
  \citenamefont {Dilhaire},\ and\ \citenamefont {Mingo}}]{pernot2010aa}%
  \BibitemOpen
  \bibfield  {author} {\bibinfo {author} {\bibfnamefont {G.}~\bibnamefont
  {Pernot}}, \bibinfo {author} {\bibfnamefont {M.}~\bibnamefont {Stoffel}},
  \bibinfo {author} {\bibfnamefont {I.}~\bibnamefont {Savic}}, \bibinfo
  {author} {\bibfnamefont {F.}~\bibnamefont {Pezzoli}}, \bibinfo {author}
  {\bibfnamefont {P.}~\bibnamefont {Chen}}, \bibinfo {author} {\bibfnamefont
  {G.}~\bibnamefont {Savelli}}, \bibinfo {author} {\bibfnamefont
  {A.}~\bibnamefont {Jacquot}}, \bibinfo {author} {\bibfnamefont
  {J.}~\bibnamefont {Schumann}}, \bibinfo {author} {\bibfnamefont
  {U.}~\bibnamefont {Denker}}, \bibinfo {author} {\bibfnamefont
  {I.}~\bibnamefont {M{\"o}nch}}, \bibinfo {author} {\bibfnamefont
  {C.}~\bibnamefont {Deneke}}, \bibinfo {author} {\bibfnamefont {O.~G.}\
  \bibnamefont {Schmidt}}, \bibinfo {author} {\bibfnamefont {J.~M.}\
  \bibnamefont {Rampnoux}}, \bibinfo {author} {\bibfnamefont {S.}~\bibnamefont
  {Wang}}, \bibinfo {author} {\bibfnamefont {M.}~\bibnamefont {Plissonnier}},
  \bibinfo {author} {\bibfnamefont {A.}~\bibnamefont {Rastelli}}, \bibinfo
  {author} {\bibfnamefont {S.}~\bibnamefont {Dilhaire}}, \ and\ \bibinfo
  {author} {\bibfnamefont {N.}~\bibnamefont {Mingo}},\ }\href
  {http://dx.doi.org/10.1038/nmat2752} {\bibfield  {journal} {\bibinfo
  {journal} {Nat Mater}\ }\textbf {\bibinfo {volume} {9}},\ \bibinfo {pages}
  {491} (\bibinfo {year} {2010})}\BibitemShut {NoStop}%
\bibitem [{\citenamefont {Losego}\ \emph {et~al.}(2013)\citenamefont {Losego},
  \citenamefont {Blitz}, \citenamefont {Vaia}, \citenamefont {Cahill},\ and\
  \citenamefont {Braun}}]{losego2013aa}%
  \BibitemOpen
  \bibfield  {author} {\bibinfo {author} {\bibfnamefont {M.~D.}\ \bibnamefont
  {Losego}}, \bibinfo {author} {\bibfnamefont {I.~P.}\ \bibnamefont {Blitz}},
  \bibinfo {author} {\bibfnamefont {R.~A.}\ \bibnamefont {Vaia}}, \bibinfo
  {author} {\bibfnamefont {D.~G.}\ \bibnamefont {Cahill}}, \ and\ \bibinfo
  {author} {\bibfnamefont {P.~V.}\ \bibnamefont {Braun}},\ }\href@noop {}
  {\bibfield  {journal} {\bibinfo  {journal} {Nano Letters}\ }\textbf {\bibinfo
  {volume} {13}},\ \bibinfo {pages} {2215} (\bibinfo {year}
  {2013})}\BibitemShut {NoStop}%
\bibitem [{\citenamefont {Hopkins}\ \emph {et~al.}(2011)\citenamefont
  {Hopkins}, \citenamefont {Mittal}, \citenamefont {Phinney}, \citenamefont
  {Grillet},\ and\ \citenamefont {Furst}}]{hopkins2011af}%
  \BibitemOpen
  \bibfield  {author} {\bibinfo {author} {\bibfnamefont {P.~E.}\ \bibnamefont
  {Hopkins}}, \bibinfo {author} {\bibfnamefont {M.}~\bibnamefont {Mittal}},
  \bibinfo {author} {\bibfnamefont {L.~M.}\ \bibnamefont {Phinney}}, \bibinfo
  {author} {\bibfnamefont {A.~M.}\ \bibnamefont {Grillet}}, \ and\ \bibinfo
  {author} {\bibfnamefont {E.~M.}\ \bibnamefont {Furst}},\ }\href {\doibase
  http://dx.doi.org/10.1063/1.3644987} {\bibfield  {journal} {\bibinfo
  {journal} {Applied Physics Letters}\ }\textbf {\bibinfo {volume} {99}},\
  \bibinfo {eid} {133106} (\bibinfo {year} {2011})}\BibitemShut {NoStop}%
\bibitem [{\citenamefont {Einstein}(1911)}]{einstein1911aa}%
  \BibitemOpen
  \bibfield  {author} {\bibinfo {author} {\bibfnamefont {A.}~\bibnamefont
  {Einstein}},\ }\href@noop {} {\bibfield  {journal} {\bibinfo  {journal}
  {Annalen der Physik}\ }\textbf {\bibinfo {volume} {340}},\ \bibinfo {pages}
  {679} (\bibinfo {year} {1911})}\BibitemShut {NoStop}%
\bibitem [{\citenamefont {S\"a\"askilahti}\ \emph {et~al.}(2014)\citenamefont
  {S\"a\"askilahti}, \citenamefont {Oksanen}, \citenamefont {Tulkki},\ and\
  \citenamefont {Volz}}]{saaskilahti2014aa}%
  \BibitemOpen
  \bibfield  {author} {\bibinfo {author} {\bibfnamefont {K.}~\bibnamefont
  {S\"a\"askilahti}}, \bibinfo {author} {\bibfnamefont {J.}~\bibnamefont
  {Oksanen}}, \bibinfo {author} {\bibfnamefont {J.}~\bibnamefont {Tulkki}}, \
  and\ \bibinfo {author} {\bibfnamefont {S.}~\bibnamefont {Volz}},\ }\href
  {\doibase 10.1103/PhysRevB.90.134312} {\bibfield  {journal} {\bibinfo
  {journal} {Phys. Rev. B}\ }\textbf {\bibinfo {volume} {90}},\ \bibinfo
  {pages} {134312} (\bibinfo {year} {2014})}\BibitemShut {NoStop}%
\bibitem [{\citenamefont {Chalopin}\ and\ \citenamefont
  {Volz}(2013)}]{chalopin2013aa}%
  \BibitemOpen
  \bibfield  {author} {\bibinfo {author} {\bibfnamefont {Y.}~\bibnamefont
  {Chalopin}}\ and\ \bibinfo {author} {\bibfnamefont {S.}~\bibnamefont
  {Volz}},\ }\href@noop {} {\bibfield  {journal} {\bibinfo  {journal} {Applied
  Physics Letters}\ }\textbf {\bibinfo {volume} {103}},\ \bibinfo {eid}
  {051602} (\bibinfo {year} {2013})}\BibitemShut {NoStop}%
\bibitem [{\citenamefont {Duda}\ \emph {et~al.}(2011)\citenamefont {Duda},
  \citenamefont {English}, \citenamefont {Piekos}, \citenamefont {Soffa},
  \citenamefont {Zhigilei},\ and\ \citenamefont {Hopkins}}]{duda2011ab}%
  \BibitemOpen
  \bibfield  {author} {\bibinfo {author} {\bibfnamefont {J.~C.}\ \bibnamefont
  {Duda}}, \bibinfo {author} {\bibfnamefont {T.~S.}\ \bibnamefont {English}},
  \bibinfo {author} {\bibfnamefont {E.~S.}\ \bibnamefont {Piekos}}, \bibinfo
  {author} {\bibfnamefont {W.~A.}\ \bibnamefont {Soffa}}, \bibinfo {author}
  {\bibfnamefont {L.~V.}\ \bibnamefont {Zhigilei}}, \ and\ \bibinfo {author}
  {\bibfnamefont {P.~E.}\ \bibnamefont {Hopkins}},\ }\href {\doibase
  10.1103/PhysRevB.84.193301} {\bibfield  {journal} {\bibinfo  {journal} {Phys.
  Rev. B}\ }\textbf {\bibinfo {volume} {84}},\ \bibinfo {pages} {193301}
  (\bibinfo {year} {2011})}\BibitemShut {NoStop}%
\bibitem [{\citenamefont {Giri}\ \emph
  {et~al.}(2016{\natexlab{a}})\citenamefont {Giri}, \citenamefont {Braun},\
  and\ \citenamefont {Hopkins}}]{giri2016af}%
  \BibitemOpen
  \bibfield  {author} {\bibinfo {author} {\bibfnamefont {A.}~\bibnamefont
  {Giri}}, \bibinfo {author} {\bibfnamefont {J.~L.}\ \bibnamefont {Braun}}, \
  and\ \bibinfo {author} {\bibfnamefont {P.~E.}\ \bibnamefont {Hopkins}},\
  }\bibfield  {booktitle} {\emph {\bibinfo {booktitle} {The Journal of Physical
  Chemistry C}},\ }\href@noop {} {\bibfield  {journal} {\bibinfo  {journal}
  {The Journal of Physical Chemistry C}\ }\textbf {\bibinfo {volume} {120}},\
  \bibinfo {pages} {24847} (\bibinfo {year} {2016}{\natexlab{a}})}\BibitemShut
  {NoStop}%
\bibitem [{\citenamefont {Gordiz}\ and\ \citenamefont
  {Henry}(2017)}]{gordiz2017aa}%
  \BibitemOpen
  \bibfield  {author} {\bibinfo {author} {\bibfnamefont {K.}~\bibnamefont
  {Gordiz}}\ and\ \bibinfo {author} {\bibfnamefont {A.}~\bibnamefont {Henry}},\
  }\href {\doibase 10.1063/1.4973573} {\bibfield  {journal} {\bibinfo
  {journal} {Journal of Applied Physics}\ }\textbf {\bibinfo {volume} {121}},\
  \bibinfo {pages} {025102} (\bibinfo {year} {2017})},\ \Eprint
  {http://arxiv.org/abs/http://dx.doi.org/10.1063/1.4973573}
  {http://dx.doi.org/10.1063/1.4973573} \BibitemShut {NoStop}%
\bibitem [{\citenamefont {Giri}\ \emph {et~al.}(2015)\citenamefont {Giri},
  \citenamefont {Hopkins}, \citenamefont {Wessel},\ and\ \citenamefont
  {Duda}}]{giri2015ad}%
  \BibitemOpen
  \bibfield  {author} {\bibinfo {author} {\bibfnamefont {A.}~\bibnamefont
  {Giri}}, \bibinfo {author} {\bibfnamefont {P.~E.}\ \bibnamefont {Hopkins}},
  \bibinfo {author} {\bibfnamefont {J.~G.}\ \bibnamefont {Wessel}}, \ and\
  \bibinfo {author} {\bibfnamefont {J.~C.}\ \bibnamefont {Duda}},\ }\href
  {\doibase http://dx.doi.org/10.1063/1.4934511} {\bibfield  {journal}
  {\bibinfo  {journal} {Journal of Applied Physics}\ }\textbf {\bibinfo
  {volume} {118}},\ \bibinfo {eid} {165303} (\bibinfo {year}
  {2015})}\BibitemShut {NoStop}%
\bibitem [{\citenamefont {Giri}\ \emph
  {et~al.}(2016{\natexlab{b}})\citenamefont {Giri}, \citenamefont {Braun},\
  and\ \citenamefont {Hopkins}}]{giri2016ac}%
  \BibitemOpen
  \bibfield  {author} {\bibinfo {author} {\bibfnamefont {A.}~\bibnamefont
  {Giri}}, \bibinfo {author} {\bibfnamefont {J.~L.}\ \bibnamefont {Braun}}, \
  and\ \bibinfo {author} {\bibfnamefont {P.~E.}\ \bibnamefont {Hopkins}},\
  }\href@noop {} {\bibfield  {journal} {\bibinfo  {journal} {Journal of Applied
  Physics}\ }\textbf {\bibinfo {volume} {119}},\ \bibinfo {eid} {235305}
  (\bibinfo {year} {2016}{\natexlab{b}})}\BibitemShut {NoStop}%
\bibitem [{\citenamefont {{Swartz}}\ and\ \citenamefont
  {{Pohl}}(1989)}]{swartz1989aa}%
  \BibitemOpen
  \bibfield  {author} {\bibinfo {author} {\bibfnamefont {E.~T.}\ \bibnamefont
  {{Swartz}}}\ and\ \bibinfo {author} {\bibfnamefont {R.~O.}\ \bibnamefont
  {{Pohl}}},\ }\href {\doibase 10.1103/RevModPhys.61.605} {\bibfield  {journal}
  {\bibinfo  {journal} {Reviews of Modern Physics}\ }\textbf {\bibinfo {volume}
  {61}},\ \bibinfo {pages} {605} (\bibinfo {year} {1989})}\BibitemShut
  {NoStop}%
\bibitem [{\citenamefont {Reddy}\ \emph {et~al.}(2005)\citenamefont {Reddy},
  \citenamefont {Castelino},\ and\ \citenamefont {Majumdar}}]{reddy2005aa}%
  \BibitemOpen
  \bibfield  {author} {\bibinfo {author} {\bibfnamefont {P.}~\bibnamefont
  {Reddy}}, \bibinfo {author} {\bibfnamefont {K.}~\bibnamefont {Castelino}}, \
  and\ \bibinfo {author} {\bibfnamefont {A.}~\bibnamefont {Majumdar}},\
  }\href@noop {} {\bibfield  {journal} {\bibinfo  {journal} {Applied Physics
  Letters}\ }\textbf {\bibinfo {volume} {87}},\ \bibinfo {eid} {211908}
  (\bibinfo {year} {2005})}\BibitemShut {NoStop}%
\bibitem [{\citenamefont {Hopkins}\ and\ \citenamefont
  {Norris}(2009)}]{hopkins2009af}%
  \BibitemOpen
  \bibfield  {author} {\bibinfo {author} {\bibfnamefont {P.~E.}\ \bibnamefont
  {Hopkins}}\ and\ \bibinfo {author} {\bibfnamefont {P.~M.}\ \bibnamefont
  {Norris}},\ }\href {http://dx.doi.org/10.1115/1.2995623} {\bibfield
  {journal} {\bibinfo  {journal} {Journal of Heat Transfer}\ }\textbf {\bibinfo
  {volume} {131}},\ \bibinfo {pages} {022402} (\bibinfo {year}
  {2009})}\BibitemShut {NoStop}%
\bibitem [{\citenamefont {Larkin}\ and\ \citenamefont
  {McGaughey}(2014)}]{larkin2014aa}%
  \BibitemOpen
  \bibfield  {author} {\bibinfo {author} {\bibfnamefont {J.~M.}\ \bibnamefont
  {Larkin}}\ and\ \bibinfo {author} {\bibfnamefont {A.~J.~H.}\ \bibnamefont
  {McGaughey}},\ }\href {\doibase 10.1103/PhysRevB.89.144303} {\bibfield
  {journal} {\bibinfo  {journal} {Phys. Rev. B}\ }\textbf {\bibinfo {volume}
  {89}},\ \bibinfo {pages} {144303} (\bibinfo {year} {2014})}\BibitemShut
  {NoStop}%
\bibitem [{\citenamefont {Braun}\ \emph {et~al.}(2016)\citenamefont {Braun},
  \citenamefont {Baker}, \citenamefont {Giri}, \citenamefont {Elahi},
  \citenamefont {Artyushkova}, \citenamefont {Beechem}, \citenamefont {Norris},
  \citenamefont {Leseman}, \citenamefont {Gaskins},\ and\ \citenamefont
  {Hopkins}}]{braun2016aa}%
  \BibitemOpen
  \bibfield  {author} {\bibinfo {author} {\bibfnamefont {J.~L.}\ \bibnamefont
  {Braun}}, \bibinfo {author} {\bibfnamefont {C.~H.}\ \bibnamefont {Baker}},
  \bibinfo {author} {\bibfnamefont {A.}~\bibnamefont {Giri}}, \bibinfo {author}
  {\bibfnamefont {M.}~\bibnamefont {Elahi}}, \bibinfo {author} {\bibfnamefont
  {K.}~\bibnamefont {Artyushkova}}, \bibinfo {author} {\bibfnamefont {T.~E.}\
  \bibnamefont {Beechem}}, \bibinfo {author} {\bibfnamefont {P.~M.}\
  \bibnamefont {Norris}}, \bibinfo {author} {\bibfnamefont {Z.~C.}\
  \bibnamefont {Leseman}}, \bibinfo {author} {\bibfnamefont {J.~T.}\
  \bibnamefont {Gaskins}}, \ and\ \bibinfo {author} {\bibfnamefont {P.~E.}\
  \bibnamefont {Hopkins}},\ }\href {\doibase 10.1103/PhysRevB.93.140201}
  {\bibfield  {journal} {\bibinfo  {journal} {Phys. Rev. B}\ }\textbf {\bibinfo
  {volume} {93}},\ \bibinfo {pages} {140201} (\bibinfo {year}
  {2016})}\BibitemShut {NoStop}%
\bibitem [{\citenamefont {Chen}\ \emph {et~al.}(2016)\citenamefont {Chen},
  \citenamefont {King}, \citenamefont {Muthuswamy}, \citenamefont {Koryttseva},
  \citenamefont {Wu},\ and\ \citenamefont {Navrotsky}}]{chen2016ab}%
  \BibitemOpen
  \bibfield  {author} {\bibinfo {author} {\bibfnamefont {J.}~\bibnamefont
  {Chen}}, \bibinfo {author} {\bibfnamefont {S.~W.}\ \bibnamefont {King}},
  \bibinfo {author} {\bibfnamefont {E.}~\bibnamefont {Muthuswamy}}, \bibinfo
  {author} {\bibfnamefont {A.}~\bibnamefont {Koryttseva}}, \bibinfo {author}
  {\bibfnamefont {D.}~\bibnamefont {Wu}}, \ and\ \bibinfo {author}
  {\bibfnamefont {A.}~\bibnamefont {Navrotsky}},\ }\href {\doibase
  10.1111/jace.14268} {\bibfield  {journal} {\bibinfo  {journal} {Journal of
  the American Ceramic Society}\ }\textbf {\bibinfo {volume} {99}},\ \bibinfo
  {pages} {2752} (\bibinfo {year} {2016})}\BibitemShut {NoStop}%
\bibitem [{\citenamefont {King}\ \emph {et~al.}(2011)\citenamefont {King},
  \citenamefont {French}, \citenamefont {Bielefeld},\ and\ \citenamefont
  {Lanford}}]{king2011aa}%
  \BibitemOpen
  \bibfield  {author} {\bibinfo {author} {\bibfnamefont {S.}~\bibnamefont
  {King}}, \bibinfo {author} {\bibfnamefont {M.}~\bibnamefont {French}},
  \bibinfo {author} {\bibfnamefont {J.}~\bibnamefont {Bielefeld}}, \ and\
  \bibinfo {author} {\bibfnamefont {W.}~\bibnamefont {Lanford}},\ }\href
  {\doibase http://dx.doi.org/10.1016/j.jnoncrysol.2011.04.001} {\bibfield
  {journal} {\bibinfo  {journal} {Journal of Non-Crystalline Solids}\ }\textbf
  {\bibinfo {volume} {357}},\ \bibinfo {pages} {2970 } (\bibinfo {year}
  {2011})}\BibitemShut {NoStop}%
\bibitem [{\citenamefont {Jeong}\ \emph {et~al.}(2012)\citenamefont {Jeong},
  \citenamefont {Zhu}, \citenamefont {Mao}, \citenamefont {Pan},\ and\
  \citenamefont {Tang}}]{jeong2012aa}%
  \BibitemOpen
  \bibfield  {author} {\bibinfo {author} {\bibfnamefont {T.}~\bibnamefont
  {Jeong}}, \bibinfo {author} {\bibfnamefont {J.-G.}\ \bibnamefont {Zhu}},
  \bibinfo {author} {\bibfnamefont {S.}~\bibnamefont {Mao}}, \bibinfo {author}
  {\bibfnamefont {T.}~\bibnamefont {Pan}}, \ and\ \bibinfo {author}
  {\bibfnamefont {Y.~J.}\ \bibnamefont {Tang}},\ }\href {\doibase
  10.1007/s10765-012-1193-1} {\bibfield  {journal} {\bibinfo  {journal}
  {International Journal of Thermophysics}\ }\textbf {\bibinfo {volume} {33}},\
  \bibinfo {pages} {1000} (\bibinfo {year} {2012})}\BibitemShut {NoStop}%
\bibitem [{\citenamefont {Hopkins}\ \emph {et~al.}(2012)\citenamefont
  {Hopkins}, \citenamefont {Baraket}, \citenamefont {Barnat}, \citenamefont
  {Beechem}, \citenamefont {Kearney}, \citenamefont {Duda}, \citenamefont
  {Robinson},\ and\ \citenamefont {Walton}}]{hopkins2012ab}%
  \BibitemOpen
  \bibfield  {author} {\bibinfo {author} {\bibfnamefont {P.~E.}\ \bibnamefont
  {Hopkins}}, \bibinfo {author} {\bibfnamefont {M.}~\bibnamefont {Baraket}},
  \bibinfo {author} {\bibfnamefont {E.~V.}\ \bibnamefont {Barnat}}, \bibinfo
  {author} {\bibfnamefont {T.~E.}\ \bibnamefont {Beechem}}, \bibinfo {author}
  {\bibfnamefont {S.~P.}\ \bibnamefont {Kearney}}, \bibinfo {author}
  {\bibfnamefont {J.~C.}\ \bibnamefont {Duda}}, \bibinfo {author}
  {\bibfnamefont {J.~T.}\ \bibnamefont {Robinson}}, \ and\ \bibinfo {author}
  {\bibfnamefont {S.~G.}\ \bibnamefont {Walton}},\ }\href@noop {} {\bibfield
  {journal} {\bibinfo  {journal} {Nano Letters}\ }\textbf {\bibinfo {volume}
  {12}},\ \bibinfo {pages} {590} (\bibinfo {year} {2012})}\BibitemShut
  {NoStop}%
\bibitem [{\citenamefont {Giri}\ \emph
  {et~al.}(2016{\natexlab{c}})\citenamefont {Giri}, \citenamefont {Niemel\"a},
  \citenamefont {Tynell}, \citenamefont {Gaskins}, \citenamefont {Donovan},
  \citenamefont {Karppinen},\ and\ \citenamefont {Hopkins}}]{giri2015ac}%
  \BibitemOpen
  \bibfield  {author} {\bibinfo {author} {\bibfnamefont {A.}~\bibnamefont
  {Giri}}, \bibinfo {author} {\bibfnamefont {J.-P.}\ \bibnamefont {Niemel\"a}},
  \bibinfo {author} {\bibfnamefont {T.}~\bibnamefont {Tynell}}, \bibinfo
  {author} {\bibfnamefont {J.~T.}\ \bibnamefont {Gaskins}}, \bibinfo {author}
  {\bibfnamefont {B.~F.}\ \bibnamefont {Donovan}}, \bibinfo {author}
  {\bibfnamefont {M.}~\bibnamefont {Karppinen}}, \ and\ \bibinfo {author}
  {\bibfnamefont {P.~E.}\ \bibnamefont {Hopkins}},\ }\href {\doibase
  10.1103/PhysRevB.93.115310} {\bibfield  {journal} {\bibinfo  {journal} {Phys.
  Rev. B}\ }\textbf {\bibinfo {volume} {93}},\ \bibinfo {pages} {115310}
  (\bibinfo {year} {2016}{\natexlab{c}})}\BibitemShut {NoStop}%
\bibitem [{\citenamefont {Fong}\ \emph {et~al.}(2016)\citenamefont {Fong},
  \citenamefont {Sood}, \citenamefont {Chen}, \citenamefont {Kumari},
  \citenamefont {Asheghi}, \citenamefont {Goodson}, \citenamefont {Gibson},\
  and\ \citenamefont {Wong}}]{fong2016aa}%
  \BibitemOpen
  \bibfield  {author} {\bibinfo {author} {\bibfnamefont {S.~W.}\ \bibnamefont
  {Fong}}, \bibinfo {author} {\bibfnamefont {A.}~\bibnamefont {Sood}}, \bibinfo
  {author} {\bibfnamefont {L.}~\bibnamefont {Chen}}, \bibinfo {author}
  {\bibfnamefont {N.}~\bibnamefont {Kumari}}, \bibinfo {author} {\bibfnamefont
  {M.}~\bibnamefont {Asheghi}}, \bibinfo {author} {\bibfnamefont {K.~E.}\
  \bibnamefont {Goodson}}, \bibinfo {author} {\bibfnamefont {G.~A.}\
  \bibnamefont {Gibson}}, \ and\ \bibinfo {author} {\bibfnamefont {H.-S.~P.}\
  \bibnamefont {Wong}},\ }\href {\doibase http://dx.doi.org/10.1063/1.4955165}
  {\bibfield  {journal} {\bibinfo  {journal} {Journal of Applied Physics}\
  }\textbf {\bibinfo {volume} {120}},\ \bibinfo {eid} {015103} (\bibinfo {year}
  {2016}),\ http://dx.doi.org/10.1063/1.4955165}\BibitemShut {NoStop}%
\bibitem [{\citenamefont {Kimling}\ \emph {et~al.}(2017)\citenamefont
  {Kimling}, \citenamefont {Philippi-Kobs}, \citenamefont {Jacobsohn},
  \citenamefont {Oepen},\ and\ \citenamefont {Cahill}}]{kimling2017aa}%
  \BibitemOpen
  \bibfield  {author} {\bibinfo {author} {\bibfnamefont {J.}~\bibnamefont
  {Kimling}}, \bibinfo {author} {\bibfnamefont {A.}~\bibnamefont
  {Philippi-Kobs}}, \bibinfo {author} {\bibfnamefont {J.}~\bibnamefont
  {Jacobsohn}}, \bibinfo {author} {\bibfnamefont {H.~P.}\ \bibnamefont
  {Oepen}}, \ and\ \bibinfo {author} {\bibfnamefont {D.~G.}\ \bibnamefont
  {Cahill}},\ }\href {\doibase 10.1103/PhysRevB.95.184305} {\bibfield
  {journal} {\bibinfo  {journal} {Phys. Rev. B}\ }\textbf {\bibinfo {volume}
  {95}},\ \bibinfo {pages} {184305} (\bibinfo {year} {2017})}\BibitemShut
  {NoStop}%
\bibitem [{\citenamefont {van Duin}\ \emph {et~al.}(2001)\citenamefont {van
  Duin}, \citenamefont {Dasgupta}, \citenamefont {Lorant},\ and\ \citenamefont
  {Goddard}}]{duin2001aa}%
  \BibitemOpen
  \bibfield  {author} {\bibinfo {author} {\bibfnamefont {A.~C.~T.}\
  \bibnamefont {van Duin}}, \bibinfo {author} {\bibfnamefont {S.}~\bibnamefont
  {Dasgupta}}, \bibinfo {author} {\bibfnamefont {F.}~\bibnamefont {Lorant}}, \
  and\ \bibinfo {author} {\bibfnamefont {W.~A.}\ \bibnamefont {Goddard}},\
  }\href {\doibase 10.1021/jp004368u} {\bibfield  {journal} {\bibinfo
  {journal} {The Journal of Physical Chemistry A}\ }\textbf {\bibinfo {volume}
  {105}},\ \bibinfo {pages} {9396} (\bibinfo {year} {2001})},\ \Eprint
  {http://arxiv.org/abs/http://dx.doi.org/10.1021/jp004368u}
  {http://dx.doi.org/10.1021/jp004368u} \BibitemShut {NoStop}%
\bibitem [{\citenamefont {Guo}\ \emph {et~al.}(2015)\citenamefont {Guo},
  \citenamefont {Zheng}, \citenamefont {King}, \citenamefont {Afanas'ev},
  \citenamefont {Baklanov}, \citenamefont {de~Marneffe}, \citenamefont
  {Nishi},\ and\ \citenamefont {Shohet}}]{guo2015aa}%
  \BibitemOpen
  \bibfield  {author} {\bibinfo {author} {\bibfnamefont {X.}~\bibnamefont
  {Guo}}, \bibinfo {author} {\bibfnamefont {H.}~\bibnamefont {Zheng}}, \bibinfo
  {author} {\bibfnamefont {S.~W.}\ \bibnamefont {King}}, \bibinfo {author}
  {\bibfnamefont {V.~V.}\ \bibnamefont {Afanas'ev}}, \bibinfo {author}
  {\bibfnamefont {M.~R.}\ \bibnamefont {Baklanov}}, \bibinfo {author}
  {\bibfnamefont {J.-F.}\ \bibnamefont {de~Marneffe}}, \bibinfo {author}
  {\bibfnamefont {Y.}~\bibnamefont {Nishi}}, \ and\ \bibinfo {author}
  {\bibfnamefont {J.~L.}\ \bibnamefont {Shohet}},\ }\href {\doibase
  10.1063/1.4929702} {\bibfield  {journal} {\bibinfo  {journal} {Applied
  Physics Letters}\ }\textbf {\bibinfo {volume} {107}},\ \bibinfo {pages}
  {082903} (\bibinfo {year} {2015})},\ \Eprint
  {http://arxiv.org/abs/http://dx.doi.org/10.1063/1.4929702}
  {http://dx.doi.org/10.1063/1.4929702} \BibitemShut {NoStop}%
\bibitem [{\citenamefont {King}\ \emph {et~al.}(2012)\citenamefont {King},
  \citenamefont {Jacob}, \citenamefont {Vanleuven}, \citenamefont {Colvin},
  \citenamefont {Kelly}, \citenamefont {French}, \citenamefont {Bielefeld},
  \citenamefont {Dutta}, \citenamefont {Liu},\ and\ \citenamefont
  {Gidley}}]{king2012ab}%
  \BibitemOpen
  \bibfield  {author} {\bibinfo {author} {\bibfnamefont {S.~W.}\ \bibnamefont
  {King}}, \bibinfo {author} {\bibfnamefont {D.}~\bibnamefont {Jacob}},
  \bibinfo {author} {\bibfnamefont {D.}~\bibnamefont {Vanleuven}}, \bibinfo
  {author} {\bibfnamefont {B.}~\bibnamefont {Colvin}}, \bibinfo {author}
  {\bibfnamefont {J.}~\bibnamefont {Kelly}}, \bibinfo {author} {\bibfnamefont
  {M.}~\bibnamefont {French}}, \bibinfo {author} {\bibfnamefont
  {J.}~\bibnamefont {Bielefeld}}, \bibinfo {author} {\bibfnamefont
  {D.}~\bibnamefont {Dutta}}, \bibinfo {author} {\bibfnamefont
  {M.}~\bibnamefont {Liu}}, \ and\ \bibinfo {author} {\bibfnamefont
  {D.}~\bibnamefont {Gidley}},\ }\href {\doibase 10.1149/2.021206jss}
  {\bibfield  {journal} {\bibinfo  {journal} {ECS Journal of Solid State
  Science and Technology}\ }\textbf {\bibinfo {volume} {1}},\ \bibinfo {pages}
  {N115} (\bibinfo {year} {2012})},\ \Eprint
  {http://arxiv.org/abs/http://jss.ecsdl.org/content/1/6/N115.full.pdf+html}
  {http://jss.ecsdl.org/content/1/6/N115.full.pdf+html} \BibitemShut {NoStop}%
\bibitem [{\citenamefont {Miikkulainen}\ \emph {et~al.}(2015)\citenamefont
  {Miikkulainen}, \citenamefont {Nilsen}, \citenamefont {Li}, \citenamefont
  {King}, \citenamefont {Laitinen}, \citenamefont {Sajavaara},\ and\
  \citenamefont {FjellvÃ*g}}]{miikkulainen2015aa}%
  \BibitemOpen
  \bibfield  {author} {\bibinfo {author} {\bibfnamefont {V.}~\bibnamefont
  {Miikkulainen}}, \bibinfo {author} {\bibfnamefont {O.}~\bibnamefont
  {Nilsen}}, \bibinfo {author} {\bibfnamefont {H.}~\bibnamefont {Li}}, \bibinfo
  {author} {\bibfnamefont {S.~W.}\ \bibnamefont {King}}, \bibinfo {author}
  {\bibfnamefont {M.}~\bibnamefont {Laitinen}}, \bibinfo {author}
  {\bibfnamefont {T.}~\bibnamefont {Sajavaara}}, \ and\ \bibinfo {author}
  {\bibfnamefont {H.}~\bibnamefont {FjellvÃ*g}},\ }\href {\doibase
  10.1116/1.4890006} {\bibfield  {journal} {\bibinfo  {journal} {Journal of
  Vacuum Science \& Technology A: Vacuum, Surfaces, and Films}\ }\textbf
  {\bibinfo {volume} {33}},\ \bibinfo {pages} {01A101} (\bibinfo {year}
  {2015})},\ \Eprint {http://arxiv.org/abs/http://dx.doi.org/10.1116/1.4890006}
  {http://dx.doi.org/10.1116/1.4890006} \BibitemShut {NoStop}%
\bibitem [{\citenamefont {Milosevic}\ and\ \citenamefont
  {King}(2012)}]{milosevic2012aa}%
  \BibitemOpen
  \bibfield  {author} {\bibinfo {author} {\bibfnamefont {M.}~\bibnamefont
  {Milosevic}}\ and\ \bibinfo {author} {\bibfnamefont {S.~W.}\ \bibnamefont
  {King}},\ }\href {\doibase 10.1063/1.4764346} {\bibfield  {journal} {\bibinfo
   {journal} {Journal of Applied Physics}\ }\textbf {\bibinfo {volume} {112}},\
  \bibinfo {pages} {093514} (\bibinfo {year} {2012})},\ \Eprint
  {http://arxiv.org/abs/http://dx.doi.org/10.1063/1.4764346}
  {http://dx.doi.org/10.1063/1.4764346} \BibitemShut {NoStop}%
\bibitem [{\citenamefont {King}\ and\ \citenamefont
  {Milosevic}(2012)}]{king2012aa}%
  \BibitemOpen
  \bibfield  {author} {\bibinfo {author} {\bibfnamefont {S.~W.}\ \bibnamefont
  {King}}\ and\ \bibinfo {author} {\bibfnamefont {M.}~\bibnamefont
  {Milosevic}},\ }\href {\doibase 10.1063/1.3700178} {\bibfield  {journal}
  {\bibinfo  {journal} {Journal of Applied Physics}\ }\textbf {\bibinfo
  {volume} {111}},\ \bibinfo {pages} {073109} (\bibinfo {year} {2012})},\
  \Eprint {http://arxiv.org/abs/http://dx.doi.org/10.1063/1.3700178}
  {http://dx.doi.org/10.1063/1.3700178} \BibitemShut {NoStop}%
\bibitem [{\citenamefont {Stan}\ \emph {et~al.}(2017)\citenamefont {Stan},
  \citenamefont {Gates}, \citenamefont {Hu}, \citenamefont {Kjoller},
  \citenamefont {Prater}, \citenamefont {Singh}, \citenamefont {Mays},\ and\
  \citenamefont {King}}]{stan2017aa}%
  \BibitemOpen
  \bibfield  {author} {\bibinfo {author} {\bibfnamefont {G.}~\bibnamefont
  {Stan}}, \bibinfo {author} {\bibfnamefont {R.~S.}\ \bibnamefont {Gates}},
  \bibinfo {author} {\bibfnamefont {Q.}~\bibnamefont {Hu}}, \bibinfo {author}
  {\bibfnamefont {K.}~\bibnamefont {Kjoller}}, \bibinfo {author} {\bibfnamefont
  {C.}~\bibnamefont {Prater}}, \bibinfo {author} {\bibfnamefont {K.~J.}\
  \bibnamefont {Singh}}, \bibinfo {author} {\bibfnamefont {E.}~\bibnamefont
  {Mays}}, \ and\ \bibinfo {author} {\bibfnamefont {S.~W.}\ \bibnamefont
  {King}},\ }in\ \href@noop {} {\emph {\bibinfo {booktitle} {Beilstein journal
  of nanotechnology}}}\ (\bibinfo {year} {2017})\BibitemShut {NoStop}%
\bibitem [{\citenamefont {Thomsen}\ \emph {et~al.}(1986)\citenamefont
  {Thomsen}, \citenamefont {Grahn}, \citenamefont {Maris},\ and\ \citenamefont
  {Tauc}}]{thomsen1986aa}%
  \BibitemOpen
  \bibfield  {author} {\bibinfo {author} {\bibfnamefont {C.}~\bibnamefont
  {Thomsen}}, \bibinfo {author} {\bibfnamefont {H.~T.}\ \bibnamefont {Grahn}},
  \bibinfo {author} {\bibfnamefont {H.~J.}\ \bibnamefont {Maris}}, \ and\
  \bibinfo {author} {\bibfnamefont {J.}~\bibnamefont {Tauc}},\ }\href {\doibase
  10.1103/PhysRevB.34.4129} {\bibfield  {journal} {\bibinfo  {journal} {Phys.
  Rev. B}\ }\textbf {\bibinfo {volume} {34}},\ \bibinfo {pages} {4129}
  (\bibinfo {year} {1986})}\BibitemShut {NoStop}%
\bibitem [{\citenamefont {Cahill}(2004)}]{cahill2004aa}%
  \BibitemOpen
  \bibfield  {author} {\bibinfo {author} {\bibfnamefont {D.~G.}\ \bibnamefont
  {Cahill}},\ }\href@noop {} {\bibfield  {journal} {\bibinfo  {journal} {Review
  of Scientific Instruments}\ }\textbf {\bibinfo {volume} {75}},\ \bibinfo
  {pages} {5119} (\bibinfo {year} {2004})}\BibitemShut {NoStop}%
\bibitem [{\citenamefont {Schmidt}\ \emph {et~al.}(2008)\citenamefont
  {Schmidt}, \citenamefont {Chen},\ and\ \citenamefont {Chen}}]{schmidt2008aa}%
  \BibitemOpen
  \bibfield  {author} {\bibinfo {author} {\bibfnamefont {A.~J.}\ \bibnamefont
  {Schmidt}}, \bibinfo {author} {\bibfnamefont {X.}~\bibnamefont {Chen}}, \
  and\ \bibinfo {author} {\bibfnamefont {G.}~\bibnamefont {Chen}},\ }\href
  {\doibase 10.1063/1.3006335} {\bibfield  {journal} {\bibinfo  {journal} {Rev
  Sci Instrum}\ }\textbf {\bibinfo {volume} {79}},\ \bibinfo {pages} {114902}
  (\bibinfo {year} {2008})}\BibitemShut {NoStop}%
\bibitem [{\citenamefont {Hopkins}\ \emph {et~al.}(2010)\citenamefont
  {Hopkins}, \citenamefont {Serrano}, \citenamefont {Phinney}, \citenamefont
  {Kearney}, \citenamefont {Grasser},\ and\ \citenamefont
  {Harris}}]{hopkins2010aa}%
  \BibitemOpen
  \bibfield  {author} {\bibinfo {author} {\bibfnamefont {P.~E.}\ \bibnamefont
  {Hopkins}}, \bibinfo {author} {\bibfnamefont {J.~R.}\ \bibnamefont
  {Serrano}}, \bibinfo {author} {\bibfnamefont {L.~M.}\ \bibnamefont
  {Phinney}}, \bibinfo {author} {\bibfnamefont {S.~P.}\ \bibnamefont
  {Kearney}}, \bibinfo {author} {\bibfnamefont {T.~W.}\ \bibnamefont
  {Grasser}}, \ and\ \bibinfo {author} {\bibfnamefont {C.~T.}\ \bibnamefont
  {Harris}},\ }\href {http://dx.doi.org/10.1115/1.4000993} {\bibfield
  {journal} {\bibinfo  {journal} {Journal of Heat Transfer}\ }\textbf {\bibinfo
  {volume} {132}},\ \bibinfo {pages} {081302} (\bibinfo {year}
  {2010})}\BibitemShut {NoStop}%
\bibitem [{\citenamefont {Giri}\ \emph
  {et~al.}(2016{\natexlab{d}})\citenamefont {Giri}, \citenamefont {Niemel\"a},
  \citenamefont {Szwejkowski}, \citenamefont {Karppinen},\ and\ \citenamefont
  {Hopkins}}]{giri2016aa}%
  \BibitemOpen
  \bibfield  {author} {\bibinfo {author} {\bibfnamefont {A.}~\bibnamefont
  {Giri}}, \bibinfo {author} {\bibfnamefont {J.-P.}\ \bibnamefont {Niemel\"a}},
  \bibinfo {author} {\bibfnamefont {C.~J.}\ \bibnamefont {Szwejkowski}},
  \bibinfo {author} {\bibfnamefont {M.}~\bibnamefont {Karppinen}}, \ and\
  \bibinfo {author} {\bibfnamefont {P.~E.}\ \bibnamefont {Hopkins}},\ }\href
  {\doibase 10.1103/PhysRevB.93.024201} {\bibfield  {journal} {\bibinfo
  {journal} {Phys. Rev. B}\ }\textbf {\bibinfo {volume} {93}},\ \bibinfo
  {pages} {024201} (\bibinfo {year} {2016}{\natexlab{d}})}\BibitemShut
  {NoStop}%
\end{thebibliography}
%

\end{document}